\documentclass[preprint,showpacs,preprintnumbers,amsmath,amssymb]{revtex4}
\input epsf \usepackage{graphicx}% Include figure files
\usepackage{dcolumn}% Align table columns on decimal point
\usepackage{bm}% bold math \bibliographystyle{unsrt}
\begin{document}

\title{Unfolding designable structures} 
\author{Cristiano L. Dias}
\email{diasc@physics.mcgill.ca} \affiliation{ Physics Department,
Rutherford Building, McGill University, 3600 rue University,
Montr\a'eal, Qu\a'ebec,  H3A 2T8 Canada }
\author{Martin Grant}
\email{grant@physics.mcgill.ca}
\affiliation{ Physics Department,
Rutherford Building, McGill University, 3600 rue University,
Montr\a'eal, Qu\a'ebec,  H3A 2T8 Canada }

\date{\today}

\begin{abstract}
Among an infinite number of possible folds, nature has chosen only
about 1000 distinct folds to form protein structures. Theoretical
studies suggest that selected folds are intrinsically more designable
than others; these selected folds are unusually stable, a property
called the designability principle. In this paper we use the 2D
hydrophobic-polar lattice model to classify structures according to
their designability, and Langevin dynamics to account for their time
evolution. We demonstrate that, among all possible folds, the more
designable ones are easier to unfold due to their large number of
surface-core bonds.
\end{abstract}


% 87.15.-v Biomolecules: structure and physical properties
% 87.15.Aa Theory and modeling; computer simulation
% 87.15.By Structure and bonding
% 87.15.He Dynamics and conformational changes

\maketitle

\section{Introduction}
%%%%%%%%%%%%%%%%%%%%%%

In the human body alone, the number of different proteins is estimated
to be in the range  of  50,000-100,000 and this number is even larger
in the biological world. However, when classified in terms of their
three dimensional structures, only 1000 families of protein folds are
expected to exist \cite{BREN97}. These structural templates for
sequences of amino acids can be explained \cite{LIHA96,LIHA98,MILL02}
in terms of minimalistic models where  the positions of amino acids
are restricted to lattice sites and the interaction energy between
residues is described by a coarse-grained model. In this minimalistic
approach, structures are classified according to their designability,
{\it i.e.}  the number of amino acid sequences they can
accommodate. While some structures are not used to describe proteins,
{\it i.e.} their designability is zero, a few structures are designed
by an enormous number of sequences and are, therefore, stable to amino
acid mutation -- a desirable and natural feature for evolution. Also,
highly designable structures emerge as being thermodynamically stable
\cite{LIHA96} and having protein-like symmetry
\cite{LIHA96,LIHA98,WANG00}.

Designability has also been shown to have dynamical
implications:~calculations suggest \cite{MELI99} that sequences of
amino acids that fold into highly designable structures, and are
thermodynamically stable, present a faster folding kinetics than
random sequences -- as expected for real proteins. Another important
dynamical aspect of proteins is their reaction to external force
fields:~in their natural environment, proteins have to cope with
forces during their activities. It might be expected that the set of
structures which constitute recurring protein folds react differently
to forces than other folds. In this paper we confirm this
expectation. We study the dependence of the phase diagram on
designability and show that \emph{for any combination of temperature
and shear, high-designable structures are the easiest structures to
unfold}. This result is a consequence of how the backbone (involving
strong covalent bonds) and weak bonds are distributed in these
structures.

This article is organized as follow: below we review the relation
between designability, thermodynamic stability and surface-to-core
bonds. Following this, the model to study unfolding is introduced,
together with the mathematical framework to characterize this
process. Results are then presented, followed by a discussion.

\section{Designability}
%%%%%%%%%%%%%%%%%%%%%%%

The goal of this section is to review the relation between
designability and thermodynamical stability for the hydrophobic-polar
(HP) model in the two-dimensional compact triangular lattice
\cite{IRBA98} which describes equilibrium structures of our protein
model in the next section. In the HP model \cite{LAUK89} a protein is
considered to be a chain made of polar (P) and hydrophobic (H) like
amino acids. Hydrophobicity is the only aspect of amino acids which is
taken into account since it is considered the main driving force for
folding \cite{KAUZ59}. In this coarse-grained approach the energy of a
sequence folded into a structure is given by the short-range contact
interaction:
\begin{equation}
\mathcal{H} = \sum_{i<j} \epsilon_{i,j} \Big[ \delta ( |\vec{r}_i -
\vec{r}_j| - \sigma) - \delta_{j-1,i} \Big]
\label{eqn:hamilton}
\end{equation}
where $\vec{r}_i$ is the position of the $i$ monomer and $\sigma$ is
distance between lattice sites. The first delta function allows only
nearest-neighbors interaction and the second delta excludes
interaction between residues which are adjacent along the
backbone. The interaction energy between monomers $i$ and $j$,
$\epsilon_{i,j}$, can assume 3 values depending on the type of
monomers bounded: H-H, H-P, P-P. These values are chosen to minimize
the Hamiltonian when H like amino acids are buried inside the protein
and P like amino acids are left on the surface. Following Li \emph{et
al} \cite{LIHA96}, we use: $\epsilon_{HH}=-2.3$, $\epsilon_{HP}=-1$
and $\epsilon_{PP}=0$. These values are given in arbitrary units.

\begin{figure}
\begin{center}
\begin{tabular}{cc}
\hspace{-0.1in}  \epsfxsize=1.60in  {\epsfbox{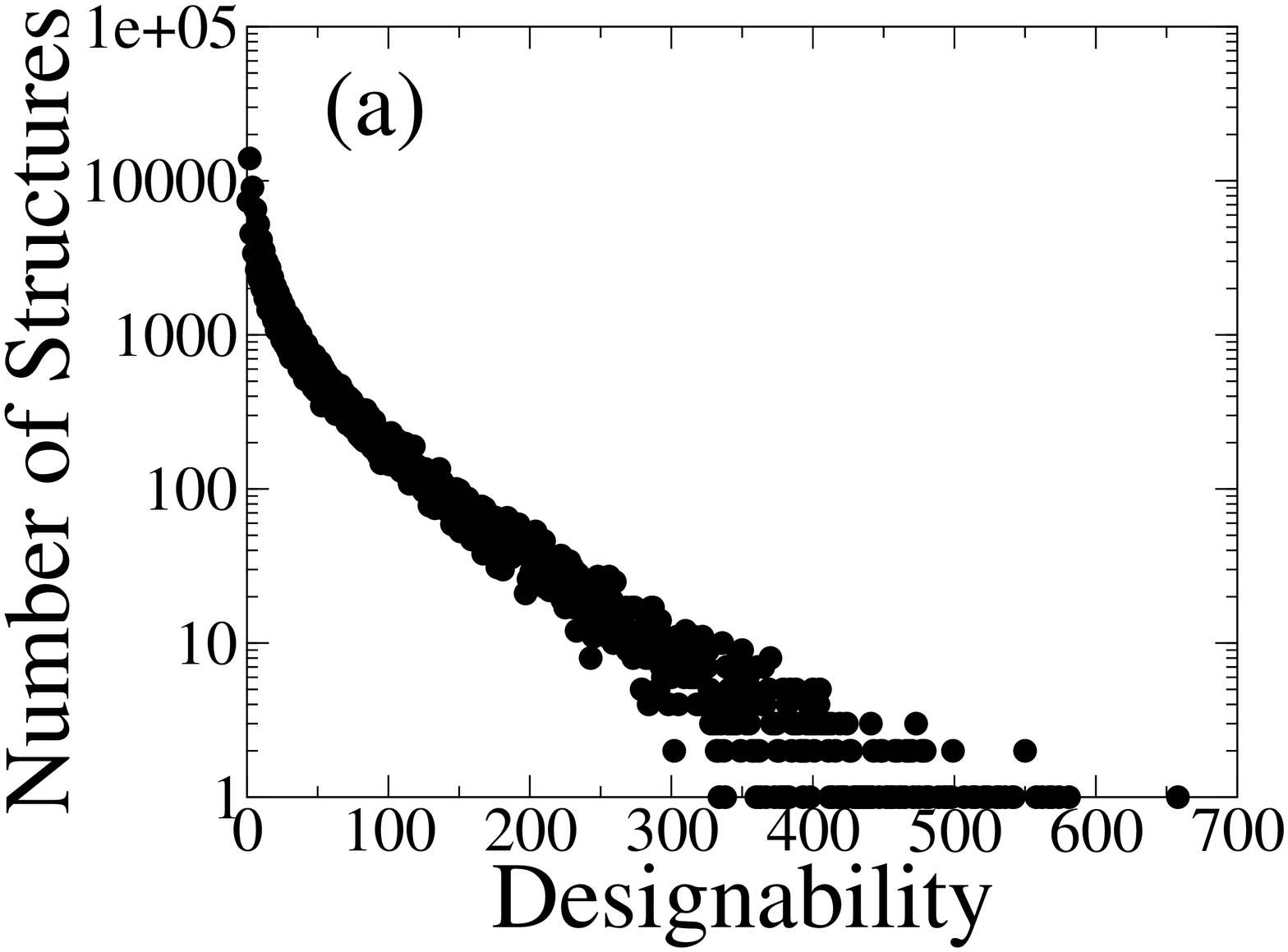}}
\vspace{0.2in} &
\hspace{0.05in} \epsfxsize=1.5in {\epsfbox{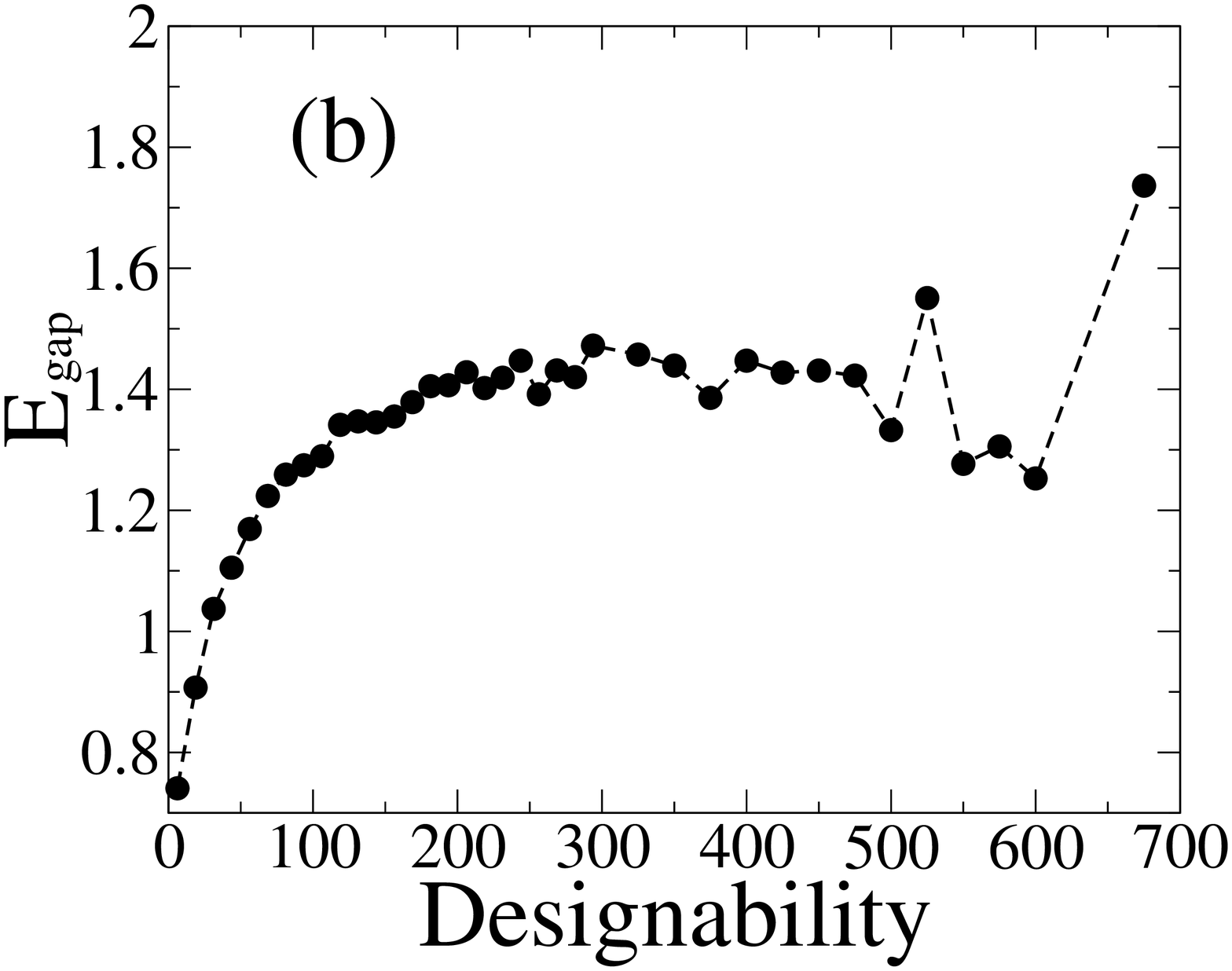}} \\
\hspace{-0.1in}\epsfxsize=1.5in  {\epsfbox{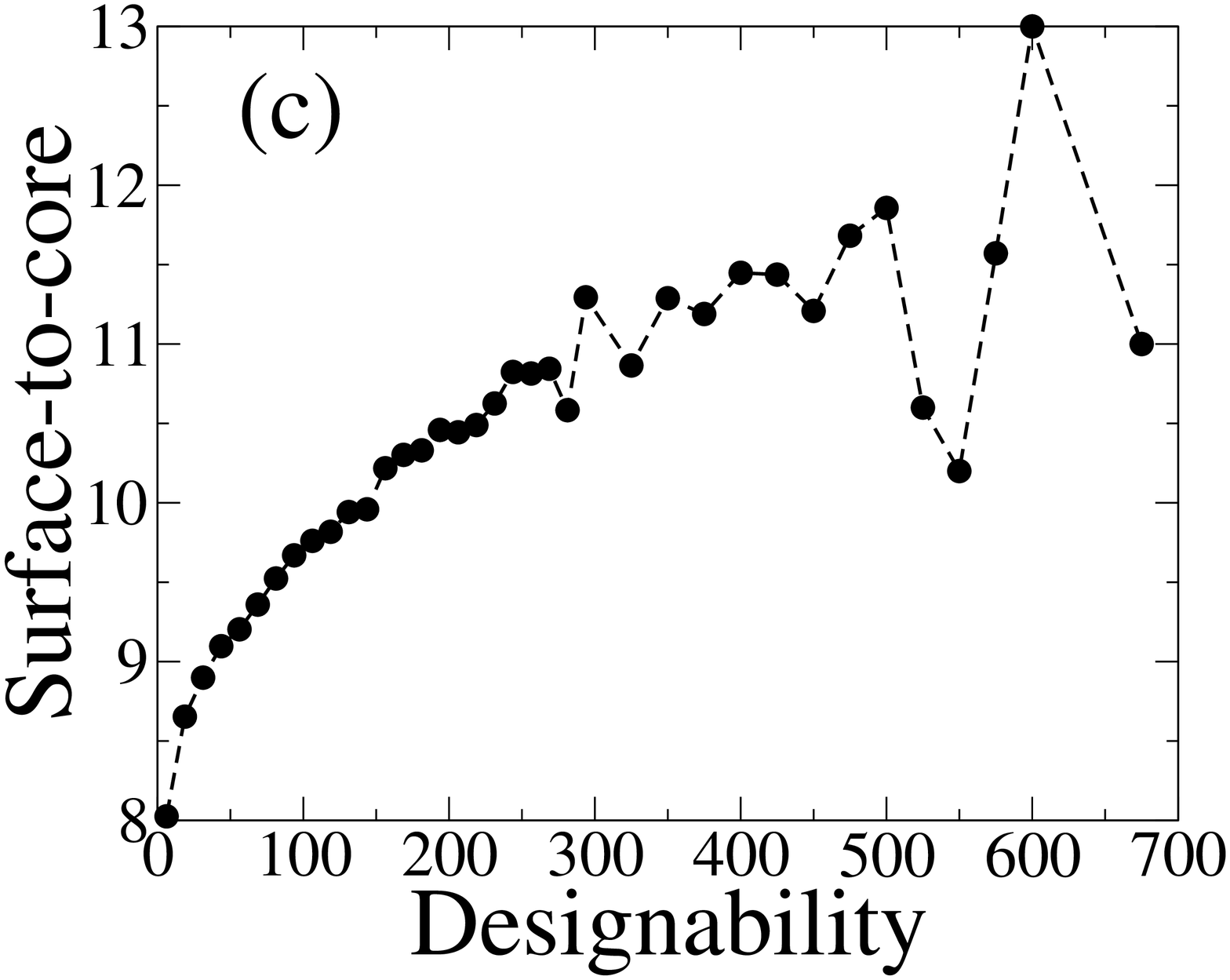}} &
\hspace{0.05in} \epsfxsize=1.5in {\epsfbox{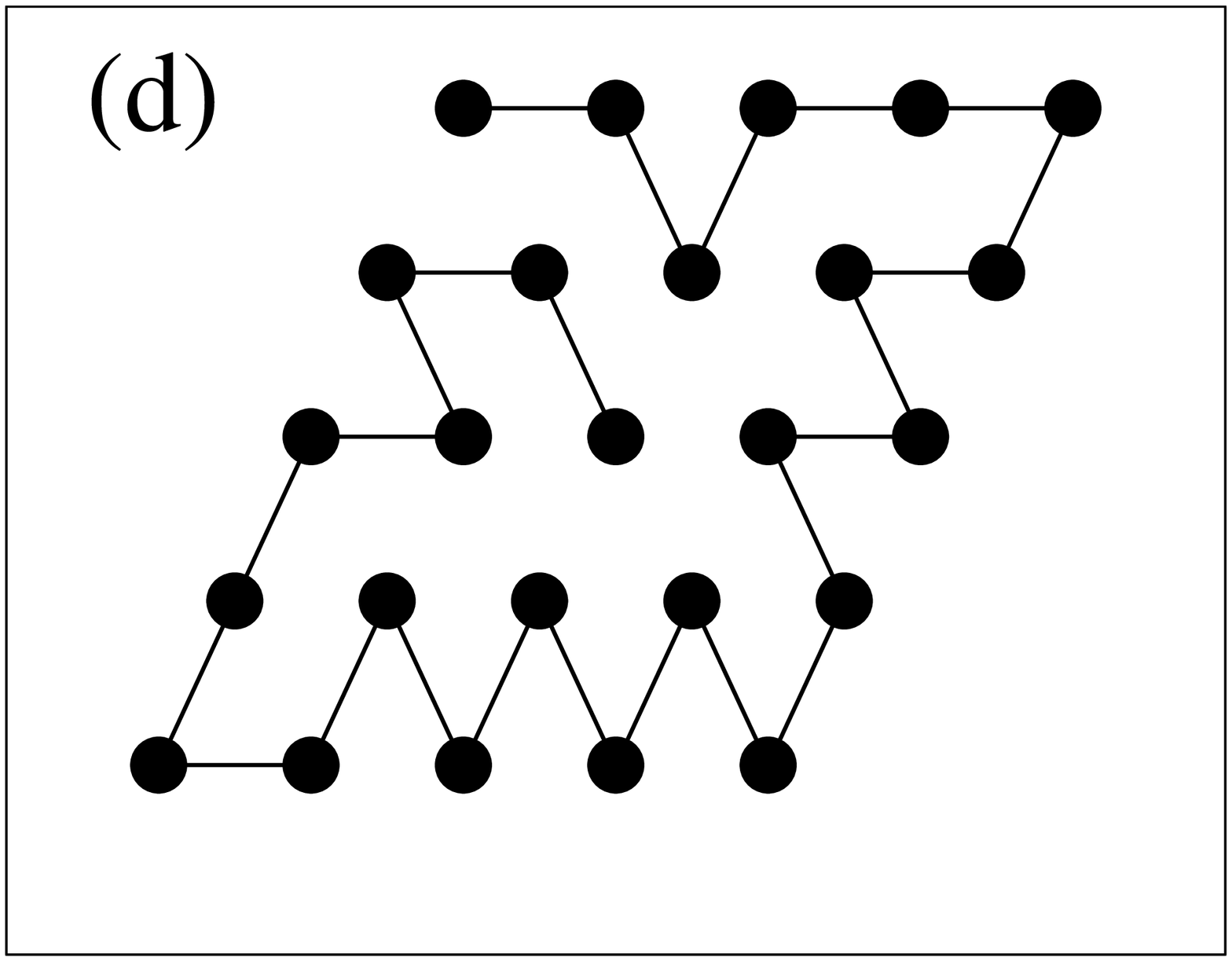}}
\end{tabular}
\end{center}
\caption{a) Histogram of designability. b) Dependence of energy gap on
designability. c) Number of bond connecting surface to core residues
versus designability. d) Fifth most designable structure.}
\label{fig:designability} 
\end{figure}

In this work, proteins are 25 amino acids long and the different
structures they can assume are restricted to compact self-avoiding
walks on a 5 X 5 triangular lattice -- see
Fig. \ref{fig:designability}(d) for an example of structure. The
number of independent structures that can be formed under these
conditions is 352,375. Now, given a sequence of amino acids, each
structure can be scanned for its native state -- \emph{i.e.} its
non-degenerate ground state. Sequences with degenerate ground states
are believed to be unrealistic since their native states are not well
defined. These sequences are therefore ignored. For our small protein,
the ground state of all its $2^{25}$ binary sequences can be computed
and we count the number of sequences that fold uniquely into a
structure. This number corresponds to the designability of the given
structure. We find that among the 352,375 structures only 135,216
($\sim 38 \%$) are non-degenerate ground states of at least one
sequence.

The distribution of designability for the 135,216 structures is given
in Fig. \ref{fig:designability}(a). Compact structures are very
different when in comes to designability: many structures have a low
value of designability, while just a rare number of folds accommodate
more than 500 sequences. These high-designable structures are on
average more stable thermodynamically than other structures. This can
be shown by computing the energy difference between the ground state
$E_o$ and the first excited state $E_1$ of a sequence:
$E_{gap}=E_1-E_o$. This energy difference is then averaged over
sequences that have the same ground state and it is a measure of the
stability of the given ground state. The correlation between $E_{gap}$
and designability is given in Fig. \ref{fig:designability}(b) where
$E_{gap}$ is averaged over a given range of designabilities.

A geometrical property of these selected structures is the large
number of bonds connecting surface monomers to core monomers
\cite{WANG00,HELL01,SHIH00}. This is illustrated in
Fig. \ref{fig:designability}(c) where the number of bonds connecting
surface to core, averaged over structures of a given range of
designability, is plotted against designability. A systematic increase
of surface-to-core bonds with designability is observed. An example of
structure which has a large number of bonds connecting surface to core
residue is the fifth most designable structure - shown in
Fig. \ref{fig:designability}(d).

\section{Model}
%%%%%%%%%%%%%%%

In the previous section we have shown that the surface of
high-designable structures is differently connected to the core of the
protein when compared to the surface of low-designable structures --
Fig. \ref{fig:designability}(c). Therefore, since unfolding starts by
unbinding surface monomers from the core, it might be expected that
the dynamics of unfolding depends on designability. To investigate
this idea, we present in this section a model to probe the dynamics of
structures in the presence of applied forces.

In this model, the energy of each structure is accounted for by two
types of potentials:~monomers which are adjacent along the backbone of
the protein interact through a harmonic potential otherwise the
interaction is via a Lennard-Jones potential. The harmonic bond
ensures that the backbone of the protein is preserved during the
simulation while monomers bound by a Lennard-Jones potential can be
driven apart, changing the structure of the protein. In this way, the
potential energy of the chain is:
\begin{equation}
V(r_{i,j}) = \sum_{i=1}^{N-1} \frac{k}{2} \big( r_{i,i+1} -
\sigma\big) ^2 + \frac{1}{2}\sum_{\stackrel{j \neq i\pm 1}{j \neq i}}
\epsilon \Big[ \Big( \frac{\sigma}{r_{i,j}} \Big) ^{12} - 2
\Big(\frac{\sigma}{r_{i,j}}\Big)^6 \Big]
\label{eqn:potenergy}
\end{equation}
in the last sum, \emph{i} and \emph{j} range from 1 to N and $r_{i,j}$
is the distance between monomers \emph{i} and \emph{j}. $\epsilon$ and
$\sigma$ are the binding energy and equilibrium length of monomers. A
cut-off distance of $2.5 \sigma$ is used for the Lennard-Jones
potential and \emph{k} is the spring constant of the harmonic
potential. Notice that the model does not discriminate between P and H
amino acids such that the dynamics of unfolding can be related
directly to the topology of the native structure independently of
amino acids sequences. Each of the 135,216 designable structures
described in the previous section corresponds to a local minima of
this potential energy and they can be viewed as \emph{equilibrium
structures}.

The fluid is modeled by including a friction and a random term
$f_i(t)$ to the force acting on each monomer (Langevin dynamics). The
intensity of the random force is given by the fluctuation-dissipation
theorem. The friction force on each monomer is proportional to the
relative velocity of the monomer with respect to the fluid: $-\gamma
\vec{v}_{rel}$ ($\gamma$ is the friction coefficient). For the
velocity of an element of the fluid located at position $\vec{r}$,
{\it i.e.} $\vec{r} = x~\hat{x} + y~\hat{y}$, we use the velocity
profile: $\vec{v}_{fluid}(\vec{r})= S y~\hat{x}$, where $S$ is the
shear rate. Inside such a fluid flow, an extended elastic object
rotates and gets stretched with an intensity dependent on its
orientation with respect to the fluid flow.

Putting the forces that act on a monomer together, its equation of
motion  inside the elongational flow is:
\begin{equation}
M \frac{d^2 \vec{r}_i}{dt^2} = \sum_{j} \vec{F}(r_{ij}) {}- M \gamma
[\dot{\vec{r}}_i - \vec{v}_{fluid}(r) ] + \vec{f}_i(t)
\label{eqn:eqnmotion}
\end{equation}
where $\vec{r}_{i}$ and $\dot{\vec{r}}_i$ are the vectors representing
the position and velocity of monomer \emph{i}. Here, $M$ is the mass
of a monomer, $\vec{F}$ is the force computed from the interacting
potential. For simplicity, $\sigma$, $\epsilon$ and $M$ are chosen to
be one. The spring is chosen to be five times stiffer than the
Lennard-Jones potential:~$k=5 (72 \epsilon / \sigma^2)$. Simulations
are carried out in units of the fastest atomic vibration:~$\tau_o = 2
\pi \sqrt{k/M}$; and the friction constant is given a value
of:~$\gamma=(\tau_o/4)^{-1}$.

\section{Results}
%%%%%%%%%%%%%%%%%%

Now we quantitatively evaluate how structures with differing
designabilities react to both thermal fluctuations and an applied
shear force. Rather than simulate all 135,216 structures, we sample as
follows. We study all the 1500 structures with highest designability,
ranging from 200 to 700. For the more numerous structures which are
less designable, we consider eight randomly-chosen structures for each
designability.  This ensemble of 3100 structures is representative of
the diversity of folds.

\subsection{Shear induced unfolding}
%%%%%%%%%%%%%%%%%%%%%%%%%%%%%%%%%%%%

Here we study how structures differing in designability react to an
applied shear force. At zero temperature, a structure only unfolds if
the shear rate is greater than $S_c$ - \emph{i.e.} when the barrier is
zero. Therefore if (at zero temperature) a structure does not unfold
at a given $S_o$ but unfolds at $S_o+\delta$, the critical shear $S_c
\equiv S_o+\delta/2$. $\delta$ being a numerical parameter accounting
for the precision of the calculation. The simulational time was 5,000
atomic vibrations. To be statistically significant we probe eight
copies of each structure to different values (all differing by $\delta
= 0.001$) of the velocity flow -- each copy having a different
orientations with respect to the fluid. Notice that $S_c$ is
proportional to the stability of a structure.

\begin{figure}
\hspace{0.5in}
\begin{center}
\epsfxsize=3in {\epsfbox{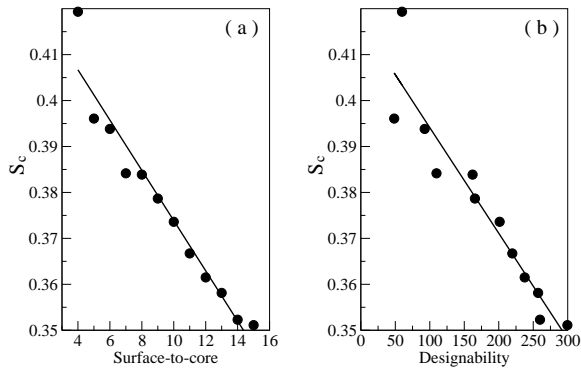}}
\end{center}
\caption{(a) Dependence of the critical velocity flow on the number of
surface-core bonds. (b) The dependence of $S_c$ on the average
designability of structures having the same number of surface-core
bonds. Lines in these figures are just a guide to the eye.}
\label{fig:mainresult} 
\end{figure}

Results are presented in Fig. \ref{fig:mainresult}(a) where the
dependence of $S_c$ on the number of times the backbone connects a
surface to a core monomer is shown. \emph{Structures with a backbone
zigzagging many times between surface and core are sensitive to small
gradients in contrast to more linear backbone structures.} In
Fig. \ref{fig:mainresult}(b) we illustrate the dependence of $S_c$ on
designability. In this Figure, both  $S_c$ and designability has been
averaged over structures having the same number of surface-to-core
bonds. \emph{A clear correlation between these quantities indicates
that structures which are highly designable require less shear to
unfold.}

\subsection{Thermal induced unfolding}
%%%%%%%%%%%%%%%%%%%%%%%%%%%%%%%%%%%%%%

\begin{figure}
\begin{center}
\vspace{0.25in} \epsfxsize=3in  {\epsfbox{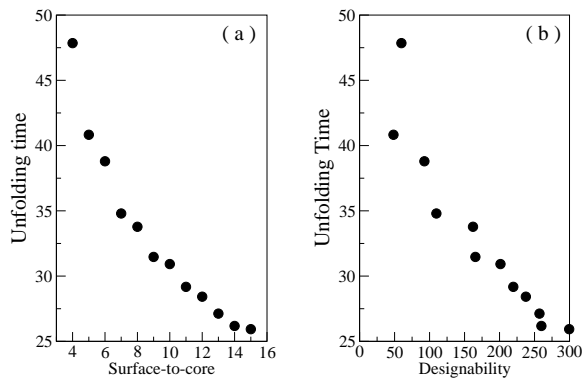}}
\end{center}
\caption{Dependence of the unfolding time on (a) number of
surface-to-core monomers and (b) designability.}
\label{fig:thermalunfolding} 
\end{figure}

In this subsection we are concerned with thermally induced
unfolding. Therefore the shear rate of our model is set to zero such
that the only cause of unfolding is thermal fluctuations.  We compute
the unfolding time of the ensemble of 3100 structures at a temperature
of 0.50 (in units of $\epsilon$). In our simulations, the unfolding
time $\tau$ is computed by tracking the population of folded
chains. The number of chains that unfold at time $t$ ($dN/dt$) is
proportional to the population of folded chains $N(t)$. In this case,
$N(t) = N_o exp(- Rt)$ where $R$ is the rate of unfolding and the
characteristic unfolding time is given by the inverse of the rate
$\tau = 1/ R$. We use 1,000 copies (\emph{i.e.} $N_o = 1000$) of each
structure in the simulations. The larger the unfolding time of a
structure, the more stable it is to thermal fluctuations. In
Fig. \ref{fig:thermalunfolding}(a) the unfolding time is plotted
versus the number of surface-to-core bonds. The clear downward trend
of this figure indicates that \emph{structures with many
surface-to-core bonds unfold faster}. Figure
\ref{fig:thermalunfolding}(b) presents the correlation between
unfolding time and designability. This Figure was obtained by
averaging both time of unfolding and designability over structures
having the same number of surface-to-core bonds. Again a clear
correlation indicates that on average, \emph{structures with low
designability are more robust to thermal fluctuations}.

\subsection{Designability dependent phase-diagram}
%%%%%%%%%%%%%%%%%%%%%%%%%%%%%%%%%%%%%%%%%%%%%%%%%%

\begin{figure}
\begin{center}
\vspace{0.25in} \epsfxsize=3in  {\epsfbox{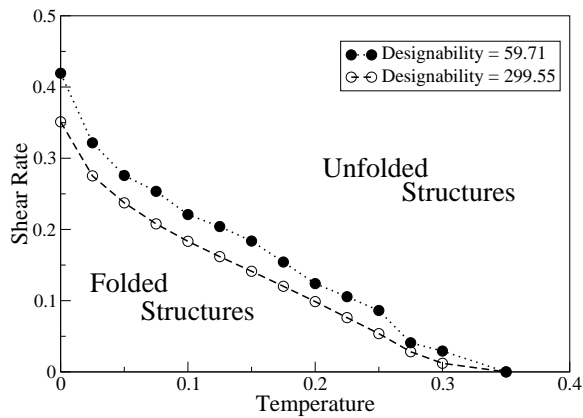}}
\end{center}
\caption{Phase diagram of the set of structures having 4 and 15
surface-to-core bonds -- filled and open circles, respectively. The
latter (former) has an average designability of 299.55 (59.71).}
\label{fig:phasediagram} 
\end{figure}

We now study how the designability of a protein affects its
phase-diagram. This diagram is constructed by computing the applied
shear rate required to unfold a structure in $5,000$ units of time at
different temperatures. This shear rate is then averaged over
structures having the same number of surface-to-core bonds. Notice
that the computed shear delimits two regions of the diagram:~folded
structures are found below this shear and unfolded structures above
it. In Fig. \ref{fig:phasediagram} the phase-diagram is shown for
structures having 4 and 15 surface-to-core bonds. These two sets of
structures have an average designability of 59.71 and 299.55
respectively. At any temperature, the set of structures with lower
designability is more robust and require a higher shear rate to
unfold. \emph{One can therefore state that high designable structures
are easier to unfold than low designable ones}.

\section{Discussion}
%%%%%%%%%%%%%%%%%%%%

The relation between thermodynamical stability and designability,
called the designability principle, has been shown in Fig.\
\ref{fig:designability}(b):~highly designable structures are more
stable thermodynamically than low designable ones. In marked contrast
to the designability principle, we have shown that highly designable
structures are easier to unfold than low designable ones --
\emph{i.e.} they are weaker. The implication is that, although highly
designable structures are more stable in the folded region of the
phase diagram, they require less perturbation to unfold. We speculate
this may be related to protein flexibility.

A qualitative explanation is as follow. We have shown that highly
designable structures are weaker due to the large number of
surface-to-core bonds they contain. Consequently, these structures
contain many small domains (\emph{i.e.} sub-structures). These are
easy to unfold:~only a few bonds need to rupture in order to separate
the domains. In contrast, low designable structures have few surface
to core bonds. As a result, many weak bonds are aligned forming
domains where monomers are correlated over long distances. For those
structures, the time of unfolding is dominated by the slow unbinding
of the largest domain. Therefore, these low designable structures can
be said to be stronger.

Also, the presence in large number of surface-to-core bonds makes it
difficult to transform highly designable structures into other
distinct compact shapes through local rearrangements of the backbone
\cite{LIHA98}. Such a transformation would require the partial
unfolding of the structure, which is unlikely in the region of the
phase diagram where folded structures are at equilibrium, followed by
folding into the new shape. Therefore, the presence of surface-to-core
bonds might explain why high designable structures are
thermodynamically stable but easier to unfolding. Finally, we expect
interesting insights to be obtained by expanding the model to three
dimensions and including hydrodynamics effects (\emph{i.e.}  modeling
the solvent explicitly).

\end{document}